%% file: separating_cd.tex
\documentclass[%
 reprint,
 superscriptaddress,
 amsmath,amssymb,
 aps,
 prx,
]{revtex4-2}

\usepackage{graphicx}
\usepackage{dcolumn}
\usepackage{bm}
\usepackage{upgreek}

\begin{document}
\include{definitions}

\title{Separating the Material and Geometry Contribution to the Circular Dichroism of Chiral Objects Made from Chiral Media}%

\author{Lukas Rebholz}%
\email{lukas.rebholz@kit.edu}%
\affiliation{Institute of Theoretical Solid State Physics, Karlsruhe Institute of Technology, Karlsruhe, Germany}%
\author{Marjan Krsti\'c}%
\affiliation{Institute of Theoretical Solid State Physics, Karlsruhe Institute of Technology, Karlsruhe, Germany}%
\author{Benedikt Zerulla}%
\affiliation{Institute of Nanotechnology, Karlsruhe Institute of Technology, Karlsruhe, Germany}%
\author{Mateusz Pawlak}%
\affiliation{Faculty of Chemistry, University of Warsaw, Warsaw, Poland}%
\author{Wiktor Lewandowski}%
\affiliation{Faculty of Chemistry, University of Warsaw, Warsaw, Poland}%
\author{Ivan Fernandez-Corbaton}%
\affiliation{Institute of Nanotechnology, Karlsruhe Institute of Technology, Karlsruhe, Germany}%
\author{Carsten Rockstuhl}%
\affiliation{Institute of Theoretical Solid State Physics, Karlsruhe Institute of Technology, Karlsruhe, Germany}%
\affiliation{Institute of Nanotechnology, Karlsruhe Institute of Technology, Karlsruhe, Germany}%

\date{\today}%

\begin{abstract}
The chirality of an object can be studied by measuring the circular dichroism, that is, the difference in absorption of light with different helicity. The chiral optical response of an object, however, can have two different origins. On the one hand, it can be linked to the chiral geometry of the object. On the other hand, it can be linked to the chiral material from which the object is made. Whereas previously, no distinction between the two contributions could be made, we report here a computational approach that allows us to separate these two contributions to the circular dichroism of an object. We consider separately the cases where geometry-related resonances affect the optical response and where they are absent. In both cases, we find the circular dichroism to be easily decomposable if a geometrically achiral object has a similar absorption spectrum to the chiral object under investigation. Furthermore, in the non-resonant case, the contribution attributed to the material can be obtained without taking any geometry into account. Besides being of fundamental importance, the possibility of disentangling both contributions will be important for guiding the future design of chiral objects and devices.
\end{abstract}

\maketitle


\section{\label{sec:introduction}Introduction}
Chiral objects exhibit a chiral optical response, i.e., they interact differently with light of different helicity \cite{schaeferling2017, wang2009}. An instance of such a phenomenon is circular dichroism (CD), which describes the difference in absorption depending on the helicity of the illumination \cite{woody1995, wang2017}. CD can be utilized to probe the chirality of a given object spectroscopically. As such, CD is at the heart of multiple applications, ranging from sensing \cite{solomon2020, warning2021} and imaging \cite{zhou2022, hu2019} to quantum devices \cite{milton2016}. To infer the properties of the examined object from such measurements, understanding the relation between CD and the chirality of the object is crucial \cite{hentschel2017, collins2017}.

For an object made from a homogeneous medium, two different causes can contribute to the chiral response. First, the chiral response can be due to the object's geometry \cite{valev2013, schaeferling2012}. A chiral response emerges if the geometry lacks mirror or inversion symmetry. This complies with the notion that a chiral object is not superimposable onto its mirror image. The second cause of chirality can be found in the properties of the medium from which the object is made \cite{mackay2009, sihvola1991}. In describing a homogeneous medium, this effect is captured by an ascribed chirality parameter in the constitutive relations representing the medium \cite{engheta1988}. Examining such a material description in detail, the cause for this material-related chirality can be understood in the same geometrical sense as before. In this case, it is the molecules or the material constituents themselves that break the mirror and inversion symmetry \cite{quack1989}. Differentiating between these two origins of chirality is a distinction concerning the length scales on which the chirality resides.

\begin{figure}[tb]
\includegraphics[width=0.48\textwidth]{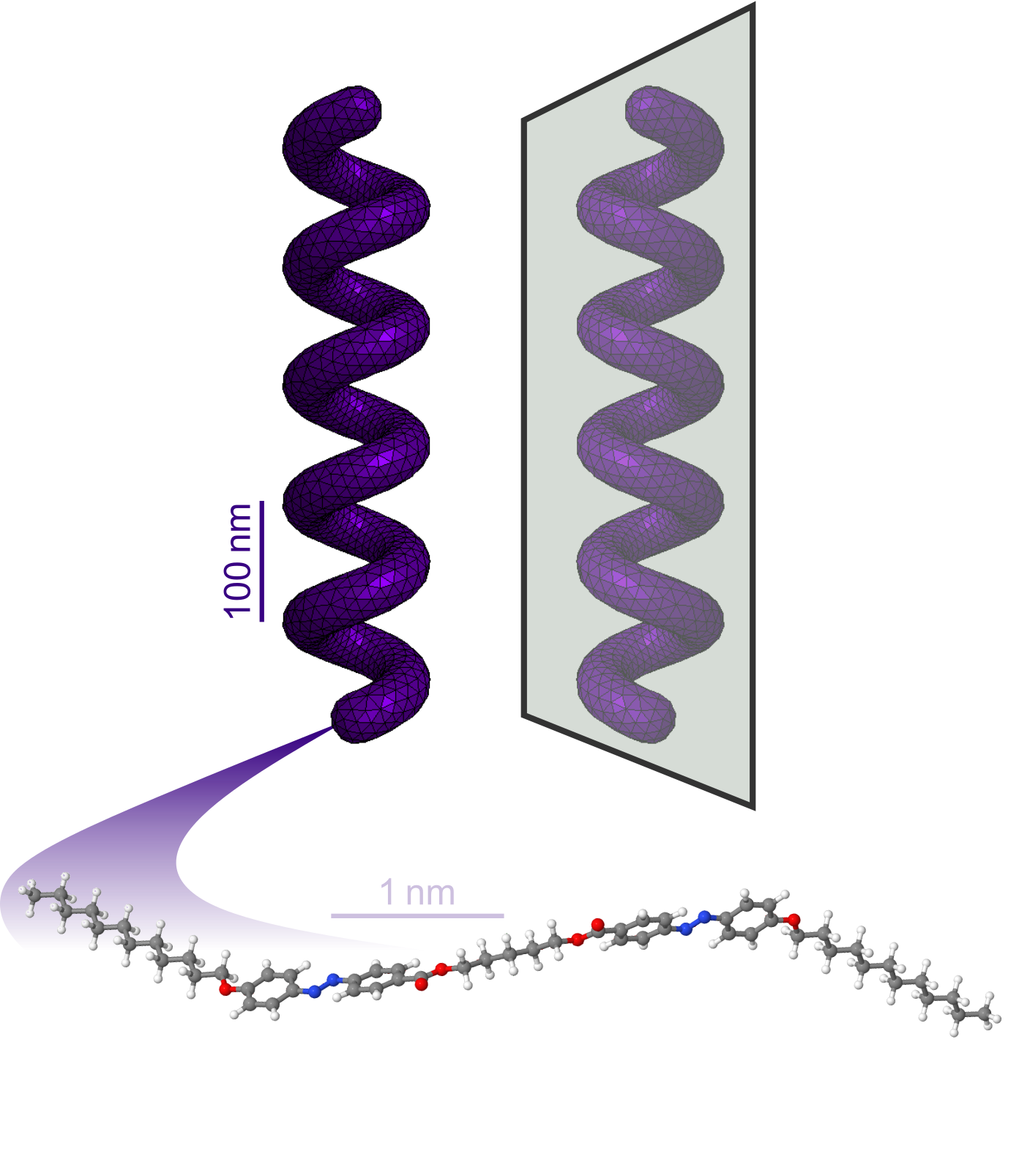}%
\caption{\label{fig:helix_mirror_schematic}Representation of an object featuring chirality via its geometric shape and material properties. The helix can be characterized by a handedness following its twist direction. Since this handedness flips upon reflection, the right-handed original helix cannot be superimposed with its left-handed mirror image. Therefore, the helix is geometrically chiral. Additionally, the helix is made from a material consisting of organic molecules that themselves are chiral. The helix, thereby, combines the geometry-related chirality with the chirality of its molecular building blocks.}
\end{figure}

For an object having both origins of chirality, these two effects always cause the overall CD. An example of such an object is shown in \fig\ref{fig:helix_mirror_schematic}. The object in the shape of a helix is geometrically chiral, which can be characterized by a handedness that describes its twist \cite{hoeflich2019, wozniak2019, lingstaedt2023}. Additionally, the object is made from a material characterized by chiral constitutive relations. In this specific case, the material is supposed to be organic and made from chiral molecules \cite{zhang2022, kotov2022}. For such a system, the question arises as to whether the CD can be dissected regarding the two origins. 

The relevance of understanding such systems becomes apparent when considering chiral organic materials that are often used for practical applications \cite{albano2020}, with the prominent examples of electroluminescent polymers \cite{yang2013}, chiral gels \cite{zhai2023}, and morphologically chiral liquid crystals \cite{park2019}. It is important to note that these materials may exhibit chirality across a range of length scales, from single molecule to supramolecular levels. Thus, beyond considering the molecular architecture of single molecules, thin films of these substances should be considered as made of elements with characteristic sizes comparable to the visible light wavelength. The precise understanding of the origin of chiroptical properties in such systems is a complex challenge \cite{wade2020}. 

Besides being of fundamental intellectual interest, separating chiral properties into individual contributions attributable to the two different scales of chirality could be of practical use as well. For example, the numerical analysis of chiral objects could be simplified by analyzing these contributions independently \cite{both2022}, choosing different, well-suited methods correspondingly. Moreover, the design of chiral objects made from a chiral material could consider both contributions as independent degrees of freedom \cite{tkachenko2014}. This would allow us to tailor chiral optical responses and widen our range of accessible dispersive chiral properties.

In the following, a computational approach is presented to explore and understand the origin of the CD of a chiral object made from a chiral molecular material. For that purpose, we combine quantum-chemical simulations \cite{casida2009}, capturing the properties of the chiral molecular material, with optical full-wave simulations \cite{lavrinenko2018}, capturing the optical response of the chiral object made from that chiral material. It is shown that the total CD can be written as a linear superposition of two contributions, each of them linked to one of the two origins: geometrical and material chirality, respectively. The helix shown in \fig\ref{fig:helix_mirror_schematic} is studied as a system in which no significant optical resonances are supported. In this case, the separation is particularly clean. Afterward, the importance of resonances is considered in studying a tetrahedral arrangement of resonant spheres. We use this to show that the distinction between the contributions remains feasible in systems where a geometrically chiral and an achiral object offer a similar optical response. This holds true in both resonant and non-resonant cases.
\begin{figure*}[tb]
\includegraphics[width=0.95\textwidth]{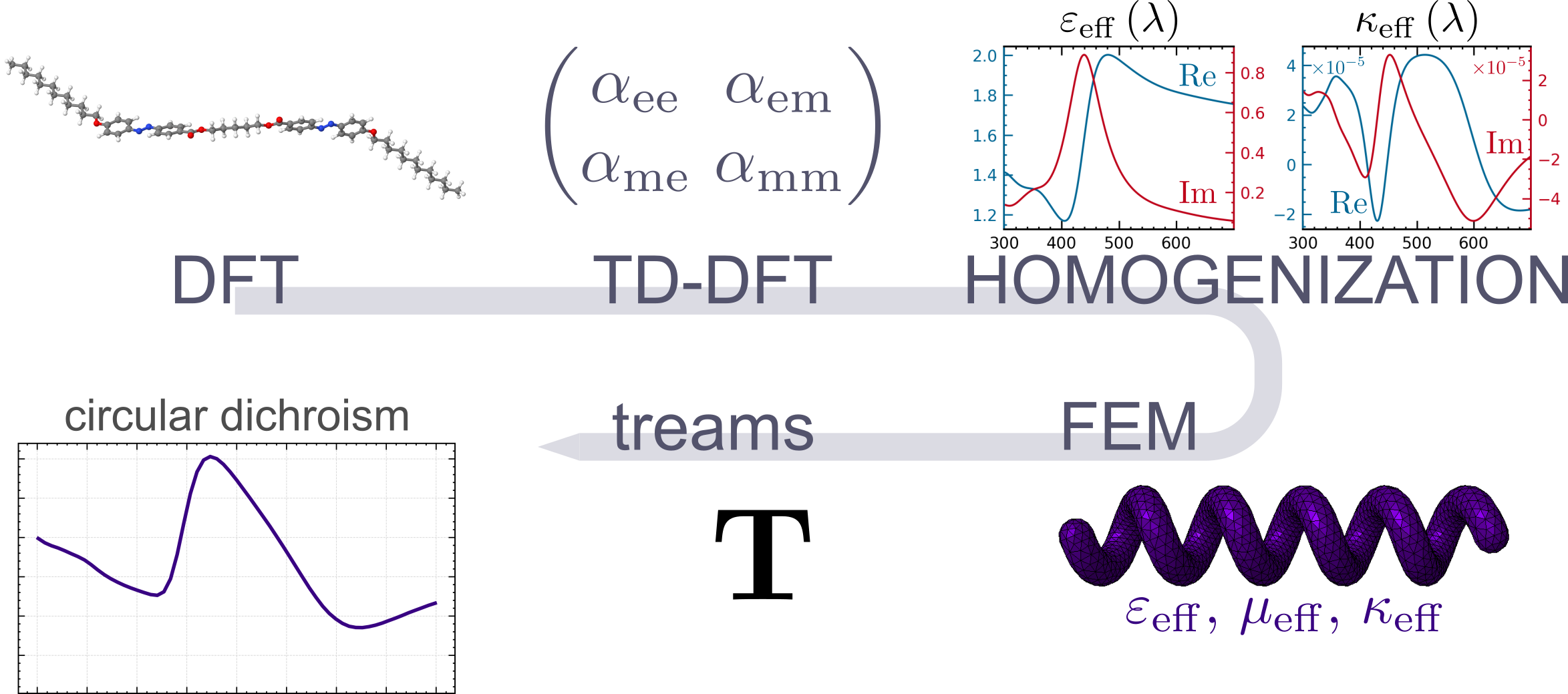}%
\caption{\label{fig:workflow}Multi-scale modeling workflow for simulating the optical response of helical nanotubes made from organic material. Starting from the structural formula of the molecule, we explore the properties of the single molecule using DFT. We arrive at the chiral molecular geometry (top left) for which the optical properties, i.e., the dynamic polarizabilities $\alpha_{ij}$, are computed using TD-DFT. According to Lorentz--Lorenz-formulas, we then find the parameters characterizing the optical properties of the molecular material $\varepsilon\subtext{eff}$, $\mu\subtext{eff}$, and $\kappa\subtext{eff}$ (top right). The material parameters describe an effective medium consisting of a specified concentration of this molecule. The effective description of the medium, together with a specification of the geometry (bottom right), is then considered in full-wave simulations based on FEM. The linear optical response of an individual helix is extracted in terms of its transition matrix (T-matrix) (bottom center). The in-house code \pkg{treams} then allows for efficient computations within the T-matrix formalism and extracting quantities of interest such as the CD (bottom left).}
\end{figure*}%

This article is structured as follows: In section \ref{sec:nonres_case}, we sketch our multi-scale approach to model the optical properties of an object, spanning different length scales. The only input we require is information about the object's geometry and the structural formula of the molecular constituents forming the material from which the object is made. The corresponding results of this approach applied to the example of an organic helix as depicted in \fig\ref{fig:helix_mirror_schematic} are discussed. Section \ref{sec:res_case} then explores the importance of resonances in this context. Afterward, we conclude with Section \ref{sec:conclusion} and give an outlook.

\section{\label{sec:nonres_case}Non-resonant case}
In \cite{jedrych2023}, the optical response of helical nanotubes forming from an organic thin film of the chiral chemical compound 12OAzo5AzoO12 is studied experimentally. Inspired by these experiments, the optical properties of organic helices are the focus of our simulations. These objects are particularly challenging as their chirality emerges at two different length scales.   

To entirely capture these aspects, we apply a multi-scale modeling approach based on \cite{Fernandez-Corbaton:2020,SURMOFCavity,zerulla2023, zerulla2023b}. The workflow of this multi-scale modeling approach is sketched in \fig\ref{fig:workflow}. Starting from the structural formula of the compound 12OAzo5AzoO12, we determine its molecular geometry from density functional theory (DFT) simulations. We found the structure of the chiral molecular geometry shown in \fig\ref{fig:workflow} at the top left; see also the section~\ref{sec:methods}~Methods in the Appendix. Afterward, time-dependent DFT (TD-DFT) as implemented in the TURBOMOLE quantum chemistry package \cite{TURBOMOLE2022, TM_today_tomorrow} is applied to calculate the electronic optical properties and dynamic polarizability tensors of a single molecule in the UV-Vis part of the spectrum. Based on the dynamic polarizability, effective material parameters that model a homogeneous, bi-isotropic medium consisting of randomly oriented molecules are derived \cite{zerulla2023}. Similarly to \cite{SURMOFCavity}, we choose a molecular concentration of $c = 3 \times 10^{26} \, \mathrm{m}^{-3}$ to arrive at values of the relative permittivity that appear reasonable for the organic compound.

Accordingly, the organic material will be represented in the following via the three complex-valued scalar parameters $\varepsilon\subtext{eff}\left( \omega \right)$, $\mu\subtext{eff}\left( \omega \right)$, and $\kappa\subtext{eff}\left( \omega \right)$. They relate the electric and magnetic field strengths and flux densities as described by the common constitutive relations \cite[Chapter 1]{kristensson2016}
\begin{align}
    \begin{pmatrix}
    \frac{1}{\varepsilon_0} \textbf{D} \left( \textbf{r}, \omega \right) \\
    c_0 \textbf{B} \left( \textbf{r}, \omega \right)
    \end{pmatrix}
    =
    \begin{pmatrix}
    \varepsilon\subtext{eff} \left( \omega \right) & \mathrm{i} \kappa\subtext{eff} \left( \omega \right) \\
    - \mathrm{i} \kappa\subtext{eff} \left( \omega \right) & \mu\subtext{eff} \left( \omega \right)
    \end{pmatrix}
    \begin{pmatrix}
    \mathbf{E} \left( \textbf{r}, \omega \right) \\
    Z_0 \mathbf{H} \left( \mathbf{r}, \omega \right)
    \end{pmatrix} \, .
\end{align}%
In the following, the frequency dependence of the material parameters is omitted for the sake of brevity. Spectra of the effective relative permittivity $\varepsilon\subtext{eff}$ and the effective material chirality parameter $\kappa\subtext{eff}$ in the explored wavelength range from $300\,\mathrm{nm}$ to $700\,\mathrm{nm}$ are shown on the top right of \fig\ref{fig:workflow} (a more detailed view is shown in \fig S8 in the Supplemental Material \cite{SI}). The effective relative permittivity shows a resonance line shape centered around $440\,\mathrm{nm}$ in its imaginary part. This feature relates to the absorption properties of the molecules and is understood on the level of molecular electronic transitions; see Figs.~S2 and S3 in the SM \cite{SI}, respectively. It leads to substantial absorption by the material in the spectra shown later. We stress that while we also consider the magnetic permeability, its difference from unity is negligible, and no notable magnetic response exists.

Upon characterizing the organic material via the effective material parameters, the optical response of an object made from such a material can be computed. Here, the geometric shape is chosen to be a helix with a circular cross-section, as shown in \fig\ref{fig:helix_mirror_schematic}. We choose the parameters specifying the geometry with a view to the experimental data in \cite{jedrych2023}. Namely, a major radius of $41.25\,\mathrm{nm}$, a minor radius of $21.25\,\mathrm{nm}$, and a pitch of $100\,\mathrm{nm}$. The number of turns is set to $5$. A corresponding mesh is depicted on the bottom right of \fig\ref{fig:workflow}. The optical response of the object is simulated using the finite element method (FEM) software package \pkg{JCMsuite} \cite{jcmsuite} and extracted as a transition matrix (T-matrix) \cite{waterman1965}. This requires simulating the optical response for a specific set of illuminations and decomposing the scattered field into vector spherical waves. The T-matrix connects the amplitudes of the vector spherical waves that expand the incident field to the amplitudes of the vector spherical waves that expand the scattered field in a matrix-vector product. Using a specific vector spherical wave as the incident field in one simulation, one column of the T-matrix can be extracted. Then, by using a different vector spherical wave in each of these full-wave simulations, the T-matrix is particularly easy to extract.

All relevant experimentally observable quantities can be predicted from the retrieved T-matrix. A prominent example is the absorption cross-section. We use the in-house code \pkg{treams} to compute these quantities of interest \cite{beutel2023treams}. The circular dichroism (CD) is then defined as
\begin{align}
    \CD = \frac{\rotavg{\xabs^+} - \rotavg{\xabs^-}}{\rotavg{\xabs^+} + \rotavg{\xabs^-}}\, ,
\end{align}
where $\rotavg{\bbullet}$ denotes the orientational average.

\begin{figure*}[tb]
\includegraphics[width=0.95\textwidth]{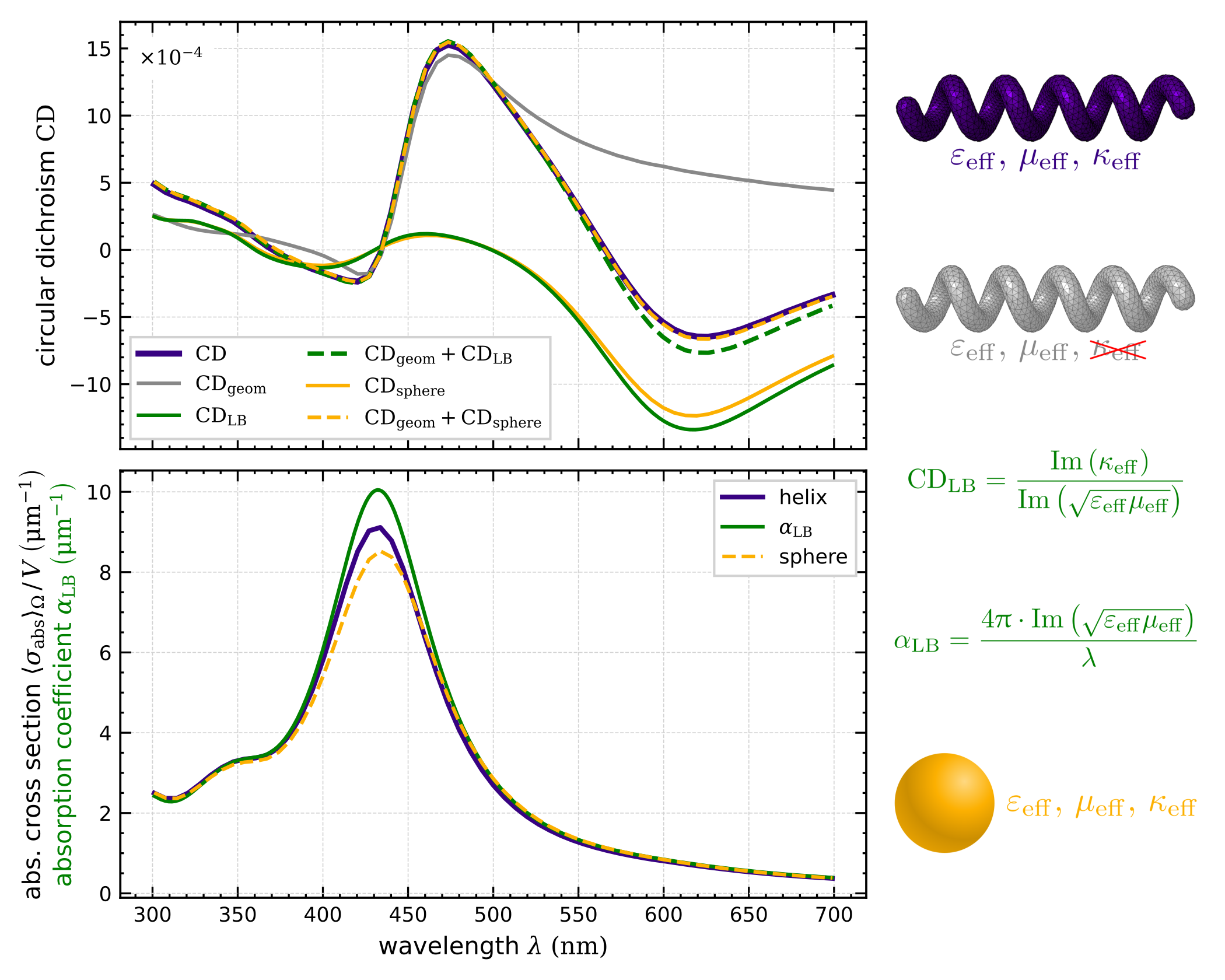}%
\caption{\label{fig:cd_xabs_nonres}Upper panel: Circular dichroism of a helix made from a chiral medium, described via effective material parameters. The total CD (dark blue) decomposes into a geometry and a material part. The geometrical CD (light gray) is found for the same helix when setting the material chirality parameter to ${\kappa\subtext{eff} = 0}$. The material-related CD matches the geometry-independent expression $\CD\subtext{LB}$ (green). The superposition (dashed green) of $\CD\subtext{LB}$ and the geometrical CD closely matches the total CD of the helix. An even better match is found from the superposition (dashed yellow) of the CD of a sphere with the same volume as the helix (yellow) and the geometrical CD. Lower panel: Absorption cross-section per unit volume and the absorption coefficient for bulk material. The absorption cross-section of the helix (dark blue) is essentially independent of whether the material chirality $\kappa\subtext{eff}$ is considered or not. A similar absorption spectrum is found for the sphere (dashed yellow). Comparison with the absorption coefficient (green) computed exclusively from the material parameters shows that the material mainly determines the absorption of the considered objects. No significant influence from geometry-related resonances is observed.}
\end{figure*}

Let us now consider the total CD signal of the helix, featuring chirality on the geometrical and molecular levels. The total CD signal, as computed from the T-matrix of the helix, is plotted in dark blue in \fig\ref{fig:cd_xabs_nonres}. The two scales of chirality of the helix are indicated by the handedness of the helix itself and by the non-zero material chirality parameter $\kappa\subtext{eff}$, respectively. 

To isolate the effect of the chiral geometry, the CD of a helix with the same geometrical features is computed for $\kappa\subtext{eff} = 0$. This corresponds to the same geometric object made from a non-chiral but absorbing material. Since $\abs{\kappa\subtext{eff}} \ll \abs{\varepsilon\subtext{eff}}$, the refractive index is not substantially changed by this.

However, the CD signal of such a helix made from an achiral material (shown as the light gray curve in \fig\ref{fig:cd_xabs_nonres}) deviates strongly from the previously considered helix made from a chiral material. By construction, this CD signal is exclusively caused by the chiral geometry of the helix and is accordingly called $\CD\subtext{geom}$ in the following. Therefore, the difference between $\CD\subtext{geom}$ and the total CD discussed before must be tied to the chirality of the material, i.e., the parameter $\kappa\subtext{eff}$. 

We want to describe this difference to the overall CD as an individual contribution originating only from the material properties. For this reason, we consider the helicity-dependent absorption found in bulk material according to the Lambert--Beer law. Here, we find the absorption coefficients
\begin{align}
    \alpha\subtext{LB}^{\pm} &= 4 \uppi \cdot \imag \left( n_{\pm} \right) \lambda^{-1}\, ,
\end{align}
where $n_{\pm} = \sqrt{\varepsilon\subtext{eff} \mu\subtext{eff}} \pm \kappa\subtext{eff}$ is the helicity-dependent refractive index. Looking at their relative difference similar to how the CD is evaluated, we define the expression
\begin{align}
    \CD\subtext{LB} &= \frac{\alpha\subtext{LB}^{+} - \alpha\subtext{LB}^{-}}{\alpha\subtext{LB}^{+} + \alpha\subtext{LB}^{-}} = \frac{\imag \left( \kappa\subtext{eff} \right)}{\imag \left( \sqrt{\varepsilon\subtext{eff} \mu\subtext{eff}} \right)} \, .
\end{align}
The subscript $\mathrm{LB}$ explicitly refers to the CD evaluated using the Lambert--Beer law for the material. This expression relates a CD signal exclusively to the effective material parameters and is entirely independent of an object's geometry. It is evaluated and presented with the solid green line in \fig\ref{fig:cd_xabs_nonres}.

The total CD is closely matched by superimposing the material contribution $\CD\subtext{LB}$ onto the isolated geometric contribution $\CD\subtext{geom}$, resulting in the dashed green line. Therefore, in this case, the CD of an object featuring chirality both in its geometry and on the molecular level can be well approximated by a decomposition into a geometry- and a material-related contribution
\begin{align}
    \CD &\approx \CD\subtext{geom} + \CD\subtext{mat} \, .
\end{align}
The fact that these show marginal interference may be explained by the different length scales on which both origins of chirality are realized. While the geometric chirality is tied to the spatial extent of the object in the order of a few hundred nanometers, the material chirality is substantiated on the molecular level. It stems from the spatially distributed electron density of the molecules. 

We have found that a geometry-independent expression describes the material-related CD contribution well. We, therefore, expect that this CD$\subtext{mat}$ also determines the CD found for objects with different shapes that are geometrically achiral and hence do not exhibit a geometric contribution to their CD. To this end, we consider the CD of a sphere, with its volume and material matching that of the helix. This is readily calculated based on analytic solutions for the optical response of a sphere and shown as the solid yellow line in \fig\ref{fig:cd_xabs_nonres}. Indeed, it closely matches the purely material-dependent CD$\subtext{LB}$. However, while the expression CD$\subtext{LB}$ is built from the properties of bulk material, introducing a spatial scale by studying an actual object with a finite volume modifies the obtained CD signal. Yet, since the geometric shape is achiral, CD$\subtext{sphere}$ is still caused by the chirality of the material.

Thus, we view the sphere’s CD as another manifestation of the material contribution to the CD and study its superposition with the helix's geometrical contribution CD$\subtext{geom}$. As seen in \fig\ref{fig:cd_xabs_nonres}, it excellently matches the total CD of the helix. In fact, this superposition with the material contribution to the CD evaluated in a sphere matches the total CD of the helix even better than for the entirely geometry-independent term CD$\subtext{LB}$. We conclude that introducing a spatial scale via a finite absorptive volume affects the material contribution to CD, and that the details of the considered geometric shape are virtually irrelevant. This makes sense since all optical resonances of the considered objects appear at much shorter wavelengths. It is then merely the finiteness of the object that affects its optical response. This depends only on the volume of the object independent of the exact shape. It is this subtle interplay between the volume of the object and its optical response that modulates the material contribution to CD.

To evaluate the influence of specifying a geometry on the absorption properties, we look at the absorption spectra of the considered objects. In the bottom panel of \fig\ref{fig:cd_xabs_nonres}, the helicity-averaged absorption cross-sections of the helix and the sphere are compared to the absorption coefficient found from the material parameters as
\begin{align}
    \alpha\subtext{LB} &= \frac{1}{2} \left( \alpha\subtext{LB}^{+} + \alpha\subtext{LB}^{-} \right) = 4 \uppi \cdot \imag \left( \sqrt{\varepsilon\subtext{eff} \mu\subtext{eff}} \right) \lambda^{-1} \, .
\end{align}
The absorption cross-sections of the helix and sphere are normalized to their respective volumes, which are identical. As can be seen from the closely matching curves, the objects’ absorption spectra are dominated by the material properties. Specifically, the resonance visible in the imaginary part of $\varepsilon\subtext{eff}$ also shows up in the absorption spectra. The line shape is slightly flattened in the absorption cross-sections of the helix and the sphere, revealing the influence of introducing a finite volume.

As a summary of these considerations, we find that the geometrical features of an object shape its absorption profile. This absorption profile, in turn, modulates the material contribution to the object's CD. This leads to the conclusion that the material contribution CD$\subtext{mat}$ of different objects only matches if they show similar absorption spectra.

For the objects examined so far, the absorption spectra were dominated by the absorption characteristic of the material parameters, and, hence, they satisfied this condition. In particular, we note that none of them support considerable geometrically induced optical resonances. However, in cases where the absorption spectra of the actual objects are significantly altered from the material's absorption characteristic, i.e., in the presence of geometrically induced resonances, the above consideration should become especially evident. In light of the above, this begs the question about the applicability of our present findings when objects that support substantial resonances are considered.

\section{\label{sec:res_case}Resonant Case}
\begin{figure*}[tb]
\includegraphics[width=0.95\textwidth]{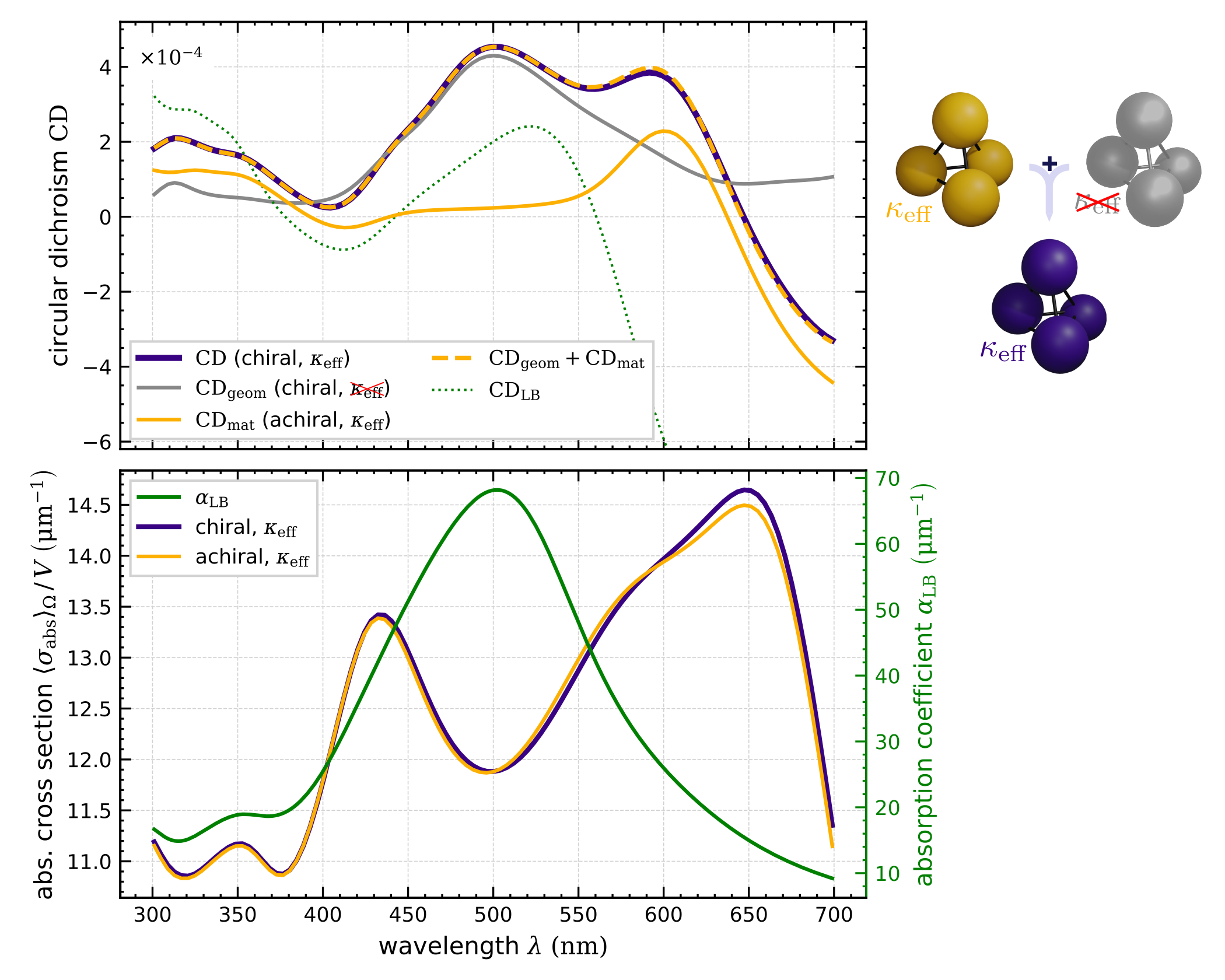}%
\caption{\label{fig:cd_xabs_res}Upper panel: Circular dichroism of tetrahedral arrangements of spheres. A geometry-related CD signal (light gray) is found for the chiral arrangement with the material chirality set to ${\kappa\subtext{eff} = 0}$. For the achiral arrangement, the material-related CD (yellow) caused by the non-zero $\kappa\subtext{eff}$ is found. Due to geometrical resonances, this $\CD\subtext{mat}$ differs from the geometry-independent term $\CD\subtext{LB}$ (dotted green). The total CD for a chiral arrangement of spheres with non-zero $\kappa\subtext{eff}$ (dark blue) is again matched by the superposition (dashed yellow) of the isolated geometry- and material-related contributions. Lower panel: Absorption cross-section per unit volume for tetrahedral arrangements of spheres. The absorption cross-sections exhibit spectral features distinctly different from the absorption coefficient for bulk material (green; right axis). These are due to geometry-related resonances. There are slight differences between the geometrically chiral (dark blue) and achiral (yellow) arrangements, underlining the geometric origin of the spectral features.}%
\end{figure*}%
To address this question, we consider in the following the CD of objects featuring geometrical and material-related chirality while supporting optical resonances. To study the isolated influence of the material chirality, we choose a tetrahedral arrangement of spheres as the starting point. This arrangement is geometrically achiral, and its non-zero CD is, therefore, tied to the chirality of the material from which the spheres are made. By changing the distances between the spheres in the tetrahedral arrangement, we arrive at a geometrically chiral configuration. As we will see below, this only causes a weak perturbation of the absorption spectrum. The geometric chirality of this arrangement, however, causes a CD that can be evaluated in isolation by considering spheres made from achiral material. The two isolated -- geometrical and material -- contributions to the CD can then be compared to the combined action of geometric and material-related chirality, which is evaluated for the chiral arrangement of spheres made from chiral material.

To make sure that the individual objects are strongly resonant, we adjust the effective material parameters. These material properties are again computed from the molecular properties of the 12OAzo5AzoO12 dimer. However, this time we consider four times the previous molecular concentration, $c = 1.2 \times 10^{27} \, \mathrm{m}^{-3}$, see \fig S9 in the SM \cite{SI}. This causes a strong dispersion in the intrinsic material parameters. The radius of the single sphere is set to $r = 80\,\mathrm{nm}$, making the single sphere resonant in the considered spectral range. The achiral tetrahedral arrangement is set up with a center-to-center distance of $220\,\mathrm{nm}$. The chiral arrangement is constructed by lengthening one of the distances between the spheres and shortening one of the distances, each by $20\,\%$. The choice of this setup is motivated by the fact that the optical response of a sphere is analytically known. The response of a coupled arrangement of spheres is readily computed using \pkg{treams}. 

To begin with, the presence of resonances and the comparability of these two arrangements are evaluated again based on the respective absorption spectra in the lower panel of \fig\ref{fig:cd_xabs_res}. The presence of resonances shows a strong modification of the absorption cross-sections compared to the absorption originating from the material parameters. Yet, both the geometrically chiral and the achiral arrangement exhibit very similar absorption spectra. Slight deviations are explained by the fact that the resonances depend on the coupling between the spheres. Slight changes in the arrangement affect, therefore, the absorption cross-section. 

The CD spectra of the different tetrahedral arrangements of spheres are plotted in the top panel of \fig\ref{fig:cd_xabs_res}. The total CD of the chiral arrangement featuring a chiral material via a non-vanishing $\kappa\subtext{eff}$ is again shown in dark blue. The geometrical contribution to the CD is computed by setting $\kappa\subtext{eff} = 0$ in this chiral arrangement. To compute the material contribution, the CD of the geometrically achiral arrangement is considered this time, with the spheres made from the chiral material. This contribution can be compared to the exclusively material-dependent term $\CD\subtext{LB}$. These two differ substantially, proving the importance of resonances for this analysis. Since resonances are not considered in $\CD\subtext{LB}$, the term is not useful for the resonant case.

Having the material- and geometry-related contributions isolated, their superposition can be used to predict the total CD with excellent accuracy. This can be clearly seen in the figure, where the dashed yellow line nearly perfectly agrees with the blue solid line. In general, slight differences can arise due to the differences between the absorption cross-sections for the geometrically chiral and achiral arrangements. It follows that even in the case where the probed objects support optical resonances, the total CD caused by the combined chirality of geometry and material separates into two independent contributions according to $\CD \approx \CD\subtext{geom} + \CD\subtext{mat}$ to an excellent approximation.

\section{\label{sec:conclusion}Conclusion}
The here applied multi-scale approach combining precise quantum chemistry simulations of optical properties of molecular materials with Maxwell solvers for photonic objects allowed us to study and understand the origin of the total CD signal for objects and devices that feature both geometry- and material-related chirality. We studied two scenarios and considered cases in which optical resonances supported by the object are present or not. In both cases, we saw that the total CD can be separated into two contributions that are attributed to the geometry- and the material-related chirality, respectively. 

For non-resonant objects, the material contribution to the CD is virtually identical and independent of the actual geometric details due to their naturally matching absorption spectra, as seen in a comparison between a helix and a sphere of the same volume. This even allowed us to analyze the material contribution without considering any geometry at all. We arrived at an expression for the material contribution to CD that depends only on the material parameters. For resonant objects, the CD remains easily decomposable in cases where an achiral object exists that features the same optical resonances as the chiral object. Therefore, in both cases, the total CD can be written as a sum of a material-dependent contribution (the CD of the geometrically achiral object made from chiral materials) and a geometry-dependent contribution (the CD of the geometrically chiral object made from achiral materials). 

The presented analysis promotes a perspective that considers geometry and material as separate originators of the chiral optical response of a given object. This could prove useful both in the computational analysis of chiral objects and in the design of devices relying on the chiral optical properties of their building blocks. 

Regarding the former, we may consider that computing the full chiral response of an object featuring material chirality in frequency-domain finite element method simulations, as used here, is computationally expensive. Comparatively, the chiral features attributed purely to the geometry are found more readily in such a setting. Capturing the effect of a non-zero material chirality parameter can render these simulations time-consuming. In contrast, the solely material-dependent CD of a sphere or the entirely geometry-independent expression CD$\subtext{LB}$ introduced in this article are easily evaluated. For a non-resonant system, the total CD could, therefore, also be analyzed by only considering the geometry in full-wave simulations and combining it with the separately and easily computed material contribution. This would save time and computational effort.

Adopting this perspective in the context of the design of optical devices means understanding geometry and material as two independent tunable elements. This offers the prospect of tailoring the chiral response of the object that is defined by the combination of those two elements. As an exemplary scenario, it would be interesting to explore the possibilities of arriving at a chosen chiral response, such as a broad, near-constant CD signal, for a given chiral material by designing the geometry accordingly. 

\begin{acknowledgments}
L.R. acknowledges support by the Karlsruhe School of Optics \& Photonics (KSOP).
L.R., I.F.C., and C.R. acknowledge support by the Deutsche Forschungsgemeinschaft (DFG, German Research Foundation) via the Collaborative Research Centre "Wave Phenomena" (SFB 1173 -- Project number 258734477).
M.K. and C.R. acknowledge support by the Deutsche Forschungsgemeinschaft (DFG, German Research Foundation) under Germany’s Excellence Strategy via the Excellence Cluster 3D Matter Made to Order (EXC-2082/1-390761711) and from the Carl Zeiss Foundation via the CZF-Focus@HEiKA Program.
M.K. and C.R. also acknowledge funding by the Volkswagen Foundation.
I.F.C. and C.R. acknowledge support by the Helmholtz Association via the Helmholtz program “Materials Systems Engineering” (MSE).
B.Z. and C.R. acknowledge support by the KIT through the “Virtual Materials Design” (VIRTMAT) project.
M.K. and C.R. acknowledge support by the state of Baden-Württemberg through bwHPC and the German Research Foundation (DFG) through grant no. INST 40/575-1 FUGG (JUSTUS 2 cluster) and the HoreKa supercomputer funded by the Ministry of Science, Research and the Arts Baden-Württemberg and by the Federal Ministry of Education and Research.
W.L. and M.P. acknowledge that this research was funded in part by the National Science Center, Poland under UMO-2020/39/O/ST5/03445 Preludium Bis grant.
We are grateful to the company JCMwave for their free provision of the FEM Maxwell solver JCMsuite.
\end{acknowledgments}

\appendix*

\section{\label{sec:methods}Methods}
The multi-scale approach employed in this work spans the scales from molecular materials to microscopic object sizes, employing the T-matrices as the bridge. T-matrices are calculated from damped dynamic polarizabilities on the quantum level using the development version of the TURBOMOLE 7.7 program package for electronic structure calculations using density functional theory \cite{TURBOMOLE2022}. Thus, we first start by looking into the structure of the single 12OAzo5AzoO12 molecule by geometry optimization in vacuum. We combined generalized-gradient approximation PBE exchange-correlation functional \cite{Perdew96a,Perdew97} with Karlsruhe def2-TZVP basis set \cite{weigendBalancedBasisSets2005, weigendAccurateCoulombfittingBasis2006} and Grimme's D4 dispersion correction \cite{caldeweyherExtensionD3Dispersion2017, caldeweyherGenerallyApplicableAtomiccharge2019} to obtain both chiral and achiral structure of the single molecular building block of the helix (Figure S1). Resolution-of-identity (RI) \cite{eichkornAuxiliaryBasisSets1995, eichkornAuxiliaryBasisSets1997} and multipole-accelerated resolution-of-identity \cite{sierkaFastEvaluationCoulomb2003} algorithms were applied to speed up the calculations without influencing the quality of the final results. To confirm that the obtained structures are the local minima on the potential energy surface of the ground state, we performed a molecular vibrational analysis. All vibrational modes had real frequencies, indicating that we found a stable configuration. Both chiral and achiral molecules had the same total energy, rendering them energetically degenerate conformers. Before tackling the calculations of the dynamic polarizabilities, we need to determine the spectral range where electronic excitations for both absorption and circular dichroism of this molecule happen. In that context, the standard approach to calculate the absorption spectrum and rotatory strengths for both conformers was applied. The complete analysis of those optical properties is given in the Supplemental Material \cite{SI}, Figures S2-S7, and it is not discussed here in detail. 

We then proceeded to calculate the electric-electric, electric-magnetic, and magnetic-magnetic dynamic polarizability tensors later used to construct T-matrices and perform homogenization of the effective material parameters. Polarizability tensors were produced for the spectral window from 300-700~nm with a wavelength resolution of 1~nm. We obtained the complex values of the polarizability elements by applying a constant damping factor of 0.1 eV for the whole spectral range. Both chiral and achiral conformers exhibited the same absorption spectrum, while only a chiral version had a CD absorption signal, as expected. The complete set of DFT and TD-DFT calculations is archived in NOMAD database under the following DOI: \href{https://doi.org/10.17172/NOMAD/2023.12.22-2}{https://doi.org/10.17172/NOMAD/2023.12.22-2}

\bibliography{separating_cd}

\end{document}

%% file: definitions.tex
\newcommand*\subtext[1]{_\text{#1}}

\newcommand*\ii{\mathrm{i}}
\newcommand*\real{\mathrm{Re}}
\newcommand*\imag{\mathrm{Im}}
\newcommand*\abs[1]{\left\lvert #1 \right\rvert}

\newcommand*\rotavg[1]{{\langle #1 \rangle}_{\Omega}}
\newcommand*\CD{\mathrm{CD}}

\newcommand*\xabs{\sigma\subtext{abs}}

\newcommand*\bbullet{\mathbf{\,\cdot\,}}

\renewcommand*\figurename{Fig.}

\newcommand*\fig{\figurename~}

\newcommand*{\pkg}[1]{\texttt{#1}}